# Polarization Transport of Transverse Acoustic Waves: Berry Phase and Spin Hall Effect of Phonons

K.Yu. Bliokh[1,2,3*] and V.D. Freilikher[3]

[1]*Institute of Radio Astronomy, 4 Krasnoznamyonnaya st., Kharkov, 61002, Ukraine*
[2]*A.Ya. Usikov Institute of Radiophysics and Electronics, 12 Akademika Proskury st., Kharkov, 61085, Ukraine*
[3]*Department of Physics, Bar-Ilan University, Ramat Gan 52900, Israel*

We carry out a detailed analysis of the short-wave (semiclassical) approximation for the linear equations of the elasticity in a smoothly inhomogeneous isotropic medium. It is shown that the polarization properties of the transverse waves are completely analogous to those of electromagnetic waves and can be considered as spin properties of optical phonons. In particular, the Hamiltonian of the transverse waves contains an additional term of the phonon spin-orbit interaction arising from the Berry gauge potential in the momentum space. This potential is diagonal in the basis of the circularly polarized waves and corresponds to the field of two 'magnetic monopoles' of opposite signs for phonons of opposite helicities. This leads to the appearance of the Berry phase in the equation for the polarization evolution and an additional "anomalous velocity" term in the ray equations. The anomalous velocity has the form of the 'Lorentz force' caused by the Berry gauge field in momentum space and gives rise to the transverse transport of waves of opposite helicities in opposite directions. This is a manifestation of the spin Hall effect of optical phonons. The effect directly relates to the conservation of total angular momentum of phonons and also influences reflection from a sharp boundary (acoustic analogue of the transverse Ferdorov–Imbert shift).

## 1. Introduction

The analogy between the linearized equations of elasticity and Maxwell equations is well known and had been pointed out in a number of textbooks (see, for instance, [1]). It helps to predict new phenomena for acoustic waves by knowing their optical counterparts. In particular, polarization phenomena in optics can be mapped onto transverse acoustic waves.

Polarization phenomena in classical electrodynamics and in elasticity theory represent collective spin properties of photons and phonons, respectively. In particular, the Berry phase for photons manifests itself as the Rytov polarization evolution law, long known in geometrical optics [2]. This law is applicable to transverse acoustic waves as well [3] (see also paper by Segert in [2]). The Berry phase arises due to a weak inter-mode coupling, which in the case of Dirac or Maxwell equations reveals itself as the *spin-orbit interaction* [4–10]. This interaction is described by an additional gauge field (the Berry gauge field) that influences the evolution of particles and waves [4–6,8–12]. Its effect on the intrinsic degrees of freedom (phase and polarization) leads to the appearance of the Berry phase, while its influence on the translational degrees of freedom (in fact, on the trajectories of particles) gives rise to the recently discovered *topological spin transport* of particles [6–12]. Examples of the topological spin transport are the anomalous and spin Hall effects in solids [12], analogous effects for relativistic electrons [6], and the optical Magnus effect [7–11]. Because of these phenomena, different polarization (spin) states of particles propagate along slightly different trajectories.

---

[*] E-mail: k_bliokh@mail.ru

In the present paper, we analyze the elasticity equations for isotropic smoothly inhomogeneous media. Based on the quantum-mechanical approach of the Berry gauge fields, we derive the modified equations of the geometric acoustics for longitudinal and transverse modes. The polarization phenomena in the propagation of the transverse acoustic waves are completely analogous to those in optics. That is, the Hamiltonian of the transverse acoustic waves contains an additional polarization term, which can be treated as the spin-orbit interaction of phonons. It is due to the non-trivial Berry gauge potential (connection) and field (curvature) in momentum space of the transverse acoustic waves. As in optics, the Berry gauge field has the form of opposite-signed 'magnetic monopoles' located at the origin of momentum space and corresponding to phonons with opposite helicities. It gives rise to the Berry phase and the Rytov polarization evolution law, and causes an additional effective force which deflects the rays in opposite directions depending on their polarization. The latter phenomenon is a manifestation of the topological spin transport or spin Hall effect of phonons. We show that the modified geometric acoustics equations obtained here are closely related to the conservation of the total angular momentum of transverse phonons and that the transverse polarization transport can also appear in the reflection or refraction on a sharp boundary.

## 2. Initial equations

The linear equations for the monochromatic wave field of displacements in an elastic inhomogeneous medium read [13]

$$\frac{\partial \sigma_{ij}}{\partial R_j} + \rho \omega^2 u_i = 0 , \qquad (1)$$

where $\mathbf{u}$ is the three-dimensional displacement vector, $\rho(\mathbf{R})$ is the density of the medium, $\mathbf{R} = (X, Y, Z)$ is the radius-vector, the summation over repeated indices is understood, and $\sigma_{ij} = a_{ijkl} \partial u_k / \partial R_l$ is the strain tensor which in an isotropic medium takes the form

$$a_{ijkl} = \lambda \delta_{ij} \delta_{kl} + \mu \left( \delta_{ik} \delta_{jl} + \delta_{il} \delta_{jk} \right) . \qquad (2)$$

Here $\lambda(\mathbf{R})$ and $\mu(\mathbf{R})$ are the Lame coefficients. Substituting Eq. (2) into Eq. (1) we obtain

$$\frac{\partial^2 u_i}{\partial R_j \partial R_j} + (m-1)\frac{\partial^2 u_j}{\partial R_i \partial R_j} + \lambdabar_{0t}^{-2} n_t^2 u_i + (m-2)\frac{\partial \ln \lambda}{\partial R_i}\frac{\partial u_j}{\partial R_j} + \frac{\partial \ln \mu}{\partial R_j}\left(\frac{\partial u_i}{\partial R_j} + \frac{\partial u_j}{\partial R_i}\right) = 0 . \qquad (3)$$

Here the following quantities are introduced:

$$c_t = \sqrt{\mu/\rho}, \ c_l = \sqrt{(\lambda+2\mu)/\rho}, \ n_{t,l} = c_{t0,l0}/c_{t,l}, \ \lambdabar_{t0,l0} = c_{t0,l0}/\omega, \ m = \left(\frac{\lambdabar_{l0} n_l}{\lambdabar_{t0} n_t}\right)^2 = \left(\frac{\lambdabar_l}{\lambdabar_t}\right)^2, \qquad (4)$$

where, $c_{t,l}$ are the local phase velocities of the transverse and longitudinal waves, $n_{t,l}$ are the local refractive indices of the transverse and longitudinal waves with respect to some 'etalon' homogeneous medium with parameters $c_{t,l} = c_{t0,l0}$, and $\lambdabar_{t0,l0}$ and $\lambdabar_{t,l}$ are the divided by $2\pi$ wavelength of the transverse and longitudinal waves in the etalon medium and in the medium under consideration, respectively.

We introduce the dimensionless differential operator of the momentum of the wave as

$$\mathbf{p} = -i\lambdabar_{t0}\frac{\partial}{\partial \mathbf{R}} . \qquad (5)$$

(In wave problems that do not contain the Planck constant explicitly, it is more convenient to use a characteristic wavelength as the "wave constant" [10]). Operator (5) obeys similar to the quantum-mechanical commutation relations

$$[R_i, p_j] = i\lambdabar_{t0}\delta_{ij} .\qquad(6)$$

Taking Eq. (5) into account, equation (3) can be written in the operator form

$$\hat{H}\mathbf{u} = 0 ,\qquad(7)$$

where the matrix operator $\hat{H}$ has the meaning of the (relativistic) Hamiltonian operator and equals

$$\hat{H}(\mathbf{p},\mathbf{R}) = p^2 - n_t^2(\mathbf{R}) + \left[m(\mathbf{R}) - 1\right]\hat{Q}(\mathbf{p}) - i\lambdabar_{t0}\hat{R}(\mathbf{p},\mathbf{R}) ,\qquad(8)$$

with

$$Q_{ij} = p_i p_j ,\quad R_{ij} = \mathbf{b}\mathbf{p} + (m-2)a_i p_j + b_j p_i ,\qquad(9)$$

where $\mathbf{a} = \nabla \ln \lambda$, $\mathbf{b} = \nabla \ln \mu$, and throughout the paper all matrix operators are marked by hats, while scalars (when they are summed up with matrices) are assumed to be multiplied by the unit matrices of the corresponding rank. Equations (7)–(9) describe the dynamics of a monochromatic field of displacements in an isotropic inhomogeneous elastic medium.

## 3. Diagonalization of elasticity equations in short-wave approximation

When the inhomogeneity of a medium is smooth, i.e. the characteristic space scale, $L$, of the variations of parameters is large compared to the wavelengths, one can use the geometric acoustics approximation (an analogue of the geometric optics or semiclassical approximations) [14], which is an asymptotic theory with respect to the small parameter

$$\varepsilon = \frac{\max(\lambdabar_{t0}, \lambdabar_{l0})}{L} \ll 1 .\qquad(10)$$

To solve equation (7)–(9) in the first, linear approximation in $\varepsilon$, we diagonalize operator $\hat{H}$, Eqs. (8) and (9), to within $\varepsilon$. It can be readily seen that the first three summands in the Hamiltonian (8) are of the order of unity (zero-order in $\varepsilon$), whereas the last summand, being proportional to the wavelength and to the gradients of the Lame coefficients, is of the order of $\varepsilon$. To diagonalize the zero-order part of Eq. (8), we note that its non-diagonal term is determined by the dyad tensor $\hat{Q} = \mathbf{p} \otimes \mathbf{p}$, Eq. (9), the same tensor that determines the non-diagonal part of the Maxwell equations [10]. Hence, in the zero approximation, operator (8) can be diagonalized by the unitary transformation similar to that for the photon electric field [10]

$$\mathbf{u} = \hat{U}(\mathbf{p})\tilde{\mathbf{u}} ,\quad \hat{U} = \begin{pmatrix} \sin\phi & \cos\theta\cos\phi & \sin\theta\cos\phi \\ -\cos\phi & \cos\theta\sin\phi & \sin\theta\sin\phi \\ 0 & -\sin\theta & \cos\theta \end{pmatrix} .\qquad(11)$$

where $(p,\theta,\phi)$ are the spherical coordinates in $\mathbf{p}$-space. Transformation (11) is a rotation, $\hat{U} \in \mathrm{SO}(3) \subset \mathrm{SU}(3)$, superposing the direction of $Z$ axis in $\mathbf{R}$-space and of the current $\mathbf{p}$ vector in $\mathbf{p}$-space (see Appendix A). Indeed, Eq. (11) leads to the transformation of the Hamiltonian, $\hat{H} \to \hat{H}' = \hat{U}^\dagger \hat{H} \hat{U}$; so that the third term in Eq. (8) becomes diagonal

$$\hat{Q}' = \hat{U}^\dagger \hat{Q} \hat{U} = \begin{pmatrix} 0 & 0 & 0 \\ 0 & 0 & 0 \\ 0 & 0 & p^2 \end{pmatrix} ,\qquad(12)$$

and coincides with Eq. (9), if $p_x = p_y = 0$ and $p_z = p$.

The diagonalization transformation does not change the first, scalar term in Eq. (8), but transforms the second one due to the non-commutativity, Eq. (6). Taking Eq. (6) into account, we get (see Appendix B):

$$\hat{U}^\dagger(\mathbf{p}) n_i^2(\mathbf{R}) \hat{U}(\mathbf{p}) = \hat{U}^\dagger(\mathbf{p}) n^2\left(i\lambdabar_{t0}\frac{\partial}{\partial \mathbf{p}}\right)\hat{U}(\mathbf{p}) = n^2\left(i\lambdabar_{t0}\frac{\partial}{\partial \mathbf{p}} + \lambdabar_{t0}\hat{\mathbf{A}}\right) = n^2\left(\mathbf{R} + \lambdabar_{t0}\hat{\mathbf{A}}\right), \quad (13)$$

where

$$\hat{\mathbf{A}}(\mathbf{p}) = i\hat{U}^\dagger \frac{\partial \hat{U}}{\partial \mathbf{p}} \quad (14)$$

is a pure gauge non-Abelian potential induced by the local gauge transformation (11) in the $\mathbf{p}$-space (see [6,10,12]), which provides the $SU(3)$ gauge invariance of the equations. Accordingly to the theory of gauge fields and to the Hamiltonian mechanics, $\mathbf{R}$ in Eq. (13) represents *canonical* (or *generalized* in the classical mechanics) coordinates, corresponding to "usual" derivatives $i\lambdabar_{t0}\partial/\partial \mathbf{p}$. At the same time, the operator $\hat{\mathbf{r}}$ is the operator of the observable (i.e. related to the center of the semiclassical particle or of the wave packet) coordinates that correspond to the *covariant* derivatives

$$\hat{\mathbf{r}} = i\lambdabar_{t0}\frac{D}{D\mathbf{p}} = i\lambdabar_{t0}\frac{\partial}{\partial \mathbf{p}} + \lambdabar_{t0}\hat{\mathbf{A}} = \mathbf{R} + \lambdabar_{t0}\hat{\mathbf{A}} . \quad (15)$$

Matrix coordinates $\hat{r}_i$ commute, $[\hat{r}_i, \hat{r}_j] = 0$, since the potential (14) is a pure gauge one and the zero field tensor corresponds to it [6,10,12]. Substitution of Eq. (11) into Eq. (14) yields

$$\hat{A}_p = 0, \quad \hat{A}_\theta = \frac{i}{p}\begin{pmatrix} 0 & 0 & 0 \\ 0 & 0 & 1 \\ 0 & -1 & 0 \end{pmatrix}, \quad \hat{A}_\phi = \frac{i}{p\sin\theta}\begin{pmatrix} 0 & -\cos\theta & -\sin\theta \\ \cos\theta & 0 & 0 \\ \sin\theta & 0 & 0 \end{pmatrix} \quad (16)$$

Potential (14), (16) is represented by means of antisymmetric hermitian matrices – generators of $SO(3)$ group (see Appendix A). Due to its non-diagonality, the second term in the Hamiltonian (8) acquires non-diagonal elements of the order of $\varepsilon$.

The last (matrix) term in the Hamiltonian (8) has no analogue in the Maxwell equations. Since it is of the order of $\varepsilon$, the contribution of commutators (6) to it is of the order of $\varepsilon^2$. Assuming that the momentum and coordinates commute, we obtain:

$$\hat{R}' = \hat{U}^\dagger \hat{R} \hat{U} \simeq \mathbf{b}\mathbf{p} + \begin{pmatrix} 0 & 0 & (m-2)\dfrac{(\mathbf{a}\times\mathbf{p})_z}{\sin\theta} \\ 0 & 0 & (m-2)\dfrac{(\mathbf{p}\times(\mathbf{a}\times\mathbf{p}))_z}{p\sin\theta} \\ \dfrac{(\mathbf{b}\times\mathbf{p})_z}{\sin\theta} & \dfrac{(\mathbf{p}\times(\mathbf{b}\times\mathbf{p}))_z}{p\sin\theta} & (m-2)\mathbf{a}\mathbf{p} + \mathbf{b}\mathbf{p} \end{pmatrix}. \quad (17)$$

Thus, after the diagonalization transformation (11) we obtain the equation $\hat{H}'\tilde{\mathbf{u}} = 0$ with the Hamiltonian

$$\hat{H}'(\mathbf{p},\mathbf{R}) = p^2 - n_i^2\left(\mathbf{R} + \lambdabar_{t0}\hat{\mathbf{A}}(\mathbf{p})\right) + \left[m(\mathbf{R}) - 1\right]\hat{Q}'(\mathbf{p}) - i\lambdabar_{t0}\hat{R}'(\mathbf{p},\mathbf{R}), \quad (18)$$

where the components are determined by Eqs. (12), (16) and (17).

In the zero approximation in $\varepsilon$ (locally homogeneous medium, or zero wavelength), the Hamiltonian (18) equals

$$\hat{H}'^{(0)}(\mathbf{p},\mathbf{R}) = p^2 - n_i^2(\mathbf{R}) + \left[m(\mathbf{R}) - 1\right]\hat{Q}'(\mathbf{p}). \quad (19)$$

This Hamiltonian is diagonal and different modes are separated. Characteristic equation for these modes is

$$\hat{H}'^{(0)}(\lambdabar_{t0}\mathbf{k}, \mathbf{R}) = 0 ,\qquad(20)$$

where $\mathbf{k}$ is the wave vector. From Eqs. (4), (12), (19) and (20) it follows that the first two levels of the Hamiltonian (19) are degenerated ($Q_{11} = Q_{22} = 0$) and correspond to waves with dispersion $\omega = kc_t$, i.e. to the transverse ('optical') shear waves with $\tilde{\mathbf{u}} = \begin{pmatrix} \tilde{u}_1 \\ \tilde{u}_2 \\ 0 \end{pmatrix}$. Similarly, the third level of the system ($Q_{33} = p^2$) corresponds to the longitudinal, compression wave with $\tilde{\mathbf{u}} = \begin{pmatrix} 0 \\ 0 \\ \tilde{u}_3 \end{pmatrix}$ and with the dispersion $\omega = kc_l$. Double degeneracy of the transverse modes is the polarization degeneracy: the transverse oscillations with different polarizations have the same dispersion in a homogeneous isotropic medium [14]. In an inhomogeneous medium, the polarization degeneracy is lifted by the non-zero gradients of the parameters [8]. The lifting of the degeneracy can be interpreted in terms of the spin-orbit interaction of phonons (see [6,7]).

In the first approximation in $\varepsilon$, the Hamiltonian (18) takes the form

$$\hat{H}'(\mathbf{p},\mathbf{R}) \simeq \hat{H}'^{(0)}(\mathbf{p},\mathbf{R}) - \lambdabar_{t0}\nabla n_t^2(\mathbf{R})\hat{\mathbf{A}}(\mathbf{p}) - i\lambdabar_{t0}\hat{R}'(\mathbf{p},\mathbf{R}) ,\qquad(21)$$

where we have expanded the second term of Eq. (18) in a Taylor series. The correction to the Hamiltonian $\hat{H}'^{(0)}$ in Eq. (21) is non-diagonal. Its upper left $2\times 2$ sector (elements with indices 11, 12, 21 and 22) contains corrections to the transverse waves, and the lower right element (with index 33) is the correction to the longitudinal wave, whereas the cross terms with indices 13, 23, 31, and 32 describe coupling and transitions between the transverse and longitudinal modes. Since the cross terms are of the order of $\varepsilon$, it follows from the adiabaticity theory that their contribution to the wave evolution is of the order of $\varepsilon^2$. Indeed, for the transverse and longitudinal waves these terms generate the appearance of longitudinal and transverse components of the field, respectively, $\tilde{u}_3 \sim \varepsilon$ and $\tilde{u}_1, \tilde{u}_2 \sim \varepsilon$ [10,14]. (The presence of such components implies minor changes in the polarization of a given mode rather than the excitation of other mode.) The $\varepsilon$-order longitudinal component contributes only to order $\varepsilon^2$ to the transverse field components and vice versa. Thus, one can neglect the cross terms in the Hamiltonian (21) [15], which leads to the breaking of gauge invariance $SU(3) \to SU(2) \times 1$. As a result, the Hamiltonian (21) and the wave equation $\hat{H}'\tilde{\mathbf{u}} = 0$ separate into two independent parts for the transverse and longitudinal waves:

$$\hat{H}^t\tilde{\mathbf{u}}_\perp = 0 ,\quad \hat{H}^t(\mathbf{p},\mathbf{R}) = p^2 - n_t^2(\mathbf{R}) - \lambdabar_{t0}\nabla n_t^2(\mathbf{R})\hat{\mathbf{A}}^t(\mathbf{p}) - i\lambdabar_{t0}\mathbf{b}(\mathbf{R})\mathbf{p} ,\qquad(22)$$

$$H^l\tilde{u}_3 = 0 ,\quad H^l(\mathbf{p},\mathbf{R}) = m(\mathbf{R})p^2 - n_l^2(\mathbf{R}) - i\lambdabar_{t0}[2\mathbf{b}(\mathbf{R})\mathbf{p} + (m-2)\mathbf{a}(\mathbf{R})\mathbf{p}] ,\qquad(23)$$

where $\tilde{\mathbf{u}}_\perp = \begin{pmatrix} \tilde{u}_1 \\ \tilde{u}_2 \end{pmatrix}$ and $\hat{A}^l \equiv A_{33} = 0$. The transverse sector of the potential (16), $\hat{\mathbf{A}}^t \equiv \begin{pmatrix} A_{11} & A_{12} \\ A_{21} & A_{22} \end{pmatrix}$, can be written (in the spherical coordinates) as:

$$\hat{\mathbf{A}}^t = p^{-1}\cot\theta(0,0,1)\hat{\sigma}_2 ,\qquad(24)$$

where $\hat{\sigma}_2 = \begin{pmatrix} 0 & -i \\ i & 0 \end{pmatrix}$ is the Pauli matrix. The components of the potential (24) commute with one another and $\hat{\mathbf{A}}^t$ is an Abelian U(1) gauge potential from $SU(2)$ sector. (The potential is Abelian due to the fact that the transverse phonon is a massless particle; in general case, it is a non-Abelian $SU(2)$ potential [6,9].) It can be transformed to a diagonal form by a global unitary transformation

$$\tilde{\mathbf{u}}_\perp = \hat{V}\psi \ , \ \ \hat{V} = \frac{1}{\sqrt{2}}\begin{pmatrix} 1 & 1 \\ i & -i \end{pmatrix}, \ \ \hat{H}^t \to V^\dagger \hat{H}^t V \ , \ \ \hat{\mathbf{A}}^t \to V^\dagger \hat{\mathbf{A}}^t V \ , \tag{25}$$

which has the meaning of the transition to the basis of circularly polarized waves (i.e., the helicity basis): $\psi = \begin{pmatrix} \psi^+ \\ \psi^- \end{pmatrix}$, $\psi^\pm = (\tilde{u}_1 \mp i\tilde{u}_2)/\sqrt{2}$ [10]. (In what follows, we use only the helicity-basis representation and notations of $\hat{H}^t$ and $\hat{\mathbf{A}}^t$ are related to this representation.) Upon transformation Eq. (25) the Hamiltonian (22) becomes diagonal (in fact, splits into two independent Hamiltonians describing the circularly polarized waves of opposite helicities):

$$\hat{H}^t \psi = 0 \ , \ \ \hat{H}^t(\mathbf{p},\mathbf{R}) = p^2 - n_t^2(\mathbf{R}) - \lambdabar_{t0} \nabla n_t^2(\mathbf{R}) \hat{\mathbf{A}}^t(\mathbf{p}) - i\lambdabar_{t0} \mathbf{b}(\mathbf{R})\mathbf{p} \ . \tag{26}$$

Here

$$\hat{\mathbf{A}}^t = p^- \cot\theta (0,0,1) \hat{\sigma}_3 \equiv \mathbf{A}^t \hat{\sigma}_3 \tag{27}$$

is a diagonal potential, and $\hat{\sigma}_3 = \begin{pmatrix} 1 & 0 \\ 0 & -1 \end{pmatrix}$. Equation (26) possesses $SU(2)$ gauge invariance which can be attributed to the spin of optical phonons. However, representation (27) shows that in fact we are dealing with single U(1) gauge potential $\mathbf{A}^t$ and U(1) gauge invariance of the equations (see Appendix A).

As seen from the Hamiltonians (22), (26), and (27), the third, proportional to the gradient of the refractive index, $\nabla n_t^2$, term in Eq. (26) lifts the degeneracy of the transverse waves. This term has the same form as that of the spin-orbit interaction of electrons and photons: it is a product of the grad of scalar potential and the Berry gauge potential [5,6,8,10]. Therefore, $\hat{H}_{SO} = -\lambdabar_{t0} \nabla n_t^2(\mathbf{R}) \hat{\mathbf{A}}^t(\mathbf{p})$ in Eq. (26) can be regarded as the *spin-orbit interaction of transverse phonons*; it couples spin (polarization) and translational degrees of freedom. Because of $\hat{H}_{SO}$, the medium can be considered as a weakly anisotropic one where the circularly polarized waves are independent normal modes, exactly as is for photons [8,10].

The operator of covariant coordinates, Eq. (15), also becomes diagonal in the present approximation. For longitudinal and circular transverse waves, respectively, it is:

$$\mathbf{r}_l = \mathbf{R} \ , \ \ \hat{\mathbf{r}}_t = \mathbf{R} + \lambdabar_{t0} \hat{\mathbf{A}}^t \ . \tag{28}$$

These are the operators of the center of wave packet for the corresponding modes. Observed coordinates of the transversely polarized wave packet can be obtained by quantum-mechanical convolution of operator (28) with the state vector (see below, Section 6).

## 4. Berry gauge field and space non-commutativity

The transverse-wave sector of the potential (14), (16), $\hat{\mathbf{A}}^t$, is no longer a pure gauge potential, since, as will be shown, a non-zero field corresponds to it. It is the Berry gauge potential, or the Berry connection, that describes the parallel transport of the vector of displacement. For the case of the transverse wave, we deal with the two-component vector $\begin{pmatrix} \tilde{u}_1 \\ \tilde{u}_2 \end{pmatrix}$ orthogonal to $\mathbf{p}$, and a natural parallel transport in the principal fiber bundle over the unit sphere of tangent vectors $\mathbf{p}/p$ in momentum space can be associated with it. This parallel transport is described by an effective vector-potential (connection) and field (curvature), generated by the 'magnetic monopole' in the origin of $\mathbf{p}$-space [2]. Indeed, the field corresponding to the potential (27) reads (in Cartesian coordinates):

$$\hat{F}^t_{ij} = \frac{\partial}{\partial p_i} \wedge \hat{A}^t_j = -e_{ijk} \frac{p_k}{p^3} \hat{\sigma}_3 \tag{29}$$

($e_{ijk}$ is the unit antisymmetric tensor). It can also be associated with the vector field $\hat{\mathbf{F}}^t$ dual to the antisymmetric tensor (29),

$$\hat{\mathbf{F}}^t = \frac{\partial}{\partial \mathbf{p}} \times \hat{\mathbf{A}}^t = -\frac{\mathbf{p}}{p^3} \hat{\sigma}_3 \equiv \mathbf{F}^t \hat{\sigma}_3 \ . \tag{29a}$$

Equations (29) and (29a) describe the Berry gauge field (Berry curvature) of the form of two 'magnetic monopoles' of opposite signs located at the origin of $\mathbf{p}$-space, which correspond to waves of opposite helicities. This Berry curvature is a particular case of the Berry gauge field for ultrarelativistic (massless) particles with well-defined helicity $\sigma$:

$$\mathbf{F}^\sigma = -\sigma \frac{\mathbf{p}}{p^3} \ . \tag{30}$$

Optical phonons, as well as photons, have helicities $\sigma = \pm 1$ related to waves of right-hand and left-hand circular polarizations. One can say that the helicity $\sigma = 0$ (prohibited for photons) corresponds to longitudinal waves whose Berry gauge field vanishes.

It is important to note that non-trivial connection and curvature in the fiber bundle over $\mathbf{p}$-space directly relate to the non-commutativity of covariant coordinates for the transverse waves [6,10,12,16,17]. Equations (28), (29), and the commutation relations (6) yield

$$[\hat{r}_{ti}, \hat{r}_{tj}] = \lambdabar_{t0} \hat{F}^t_{ij} \neq 0 \ . \tag{31}$$

As in the photon case [16], the non-commutativity of the coordinates can be attributed to the fact that a phonon in a helicity state cannot be localized. Although in the semiclassical approximation we deal with a specific polarization in the center of the wave packet, this is not a pure polarized state of the whole packet. Even if the center of the wave packet possesses pure circular polarization, the edges of the packet will be elliptically polarized due to the orthogonality condition (see [18]).

## 5. Evolution of longitudinal waves

Consider the evolution of longitudinal waves described by the Hamiltonian (23). We first rewrite Eq. (23) using Eq. (4) as

$$H^l \tilde{u}_3 = 0 \ , \quad H^l(\mathbf{p}_l, \mathbf{R}) = p_l^2 - n_l^2(\mathbf{R}) - i\lambdabar_{l0} \mathbf{p}_l \nabla \ln[\lambda(\mathbf{R}) + 2\mu(\mathbf{R})] \ , \tag{23a}$$

where, similarly to Eq. (5), we have introduced the differential operator of the momentum normalized by the longitudinal wavelength: $\mathbf{p}_l = -i\lambdabar_{l0} \partial / \partial \mathbf{R}$. The geometrical optics (acoustics) ansatz $\tilde{u}_3 = A_l(\mathbf{R}) \exp[i \lambdabar_{l0}^{-1} \Phi_l(\mathbf{R})]$ in Eq. (23a), in the zero and first approximations in $\varepsilon$ (i.e. in $\lambdabar_{l0}$) yields, respectively [3,14]:

$$(\nabla \Phi_l)^2 - n_l^2 = 0 \ , \tag{32}$$

$$2\nabla \Phi_l \nabla A_l + [\nabla^2 \Phi_l + \nabla \Phi_l \nabla \ln(\lambda + 2\mu)] A_l = 0 \ . \tag{33}$$

Equation (32) is the eikonal equation for $\Phi_l$, whereas Eq. (33) is the transport equation for the amplitude $A_l$.

We introduce the local wave vector $\mathbf{k}_l = \lambdabar_{l0}^{-1} \nabla \Phi_l$ and, corresponding to it, the dimensionless momentum $p_l = \nabla \Phi_l = \lambdabar_{l0} \mathbf{k}_l$. The transition $\mathbf{p} \to p$ corresponds to the transition from the differential momentum operator to the 'classical' momentum of the plane wave. The eikonal equation gives the dispersion equation for the longitudinal waves, $p_l = n_l$. Taking it and Eq. (4) into account, the transport equation (33) can be integrated resulting in the continuity equation

$$\nabla \left( \rho A_l^2 \mathbf{c}_l \right) = 0 \ . \tag{34}$$

Here $\mathbf{c}_l = c_l \mathbf{p}_l / p_l$ is the local phase-velocity vector. Equation (34) ensures the conservation of energy flux in the beam tube [3,14].

The rays are the real parts of the characteristics of the wave equation, Eq. (23a). The last, imaginary term in the Hamiltonian (23a) does not contribute to the real rays (it accounts only for variations of the amplitude $A_l$); therefore, they coincide with characteristics of the eikonal equation (32). The latter are described by the Hamiltonian equations with the Hamiltonian

$$\mathcal{H}^l \left( \mathbf{p}_l, \mathbf{r}_l \right) = \frac{1}{2} \left[ p_l^2 - n_l^2 \left( \mathbf{r}_l \right) \right] = 0 \ , \tag{35}$$

where the coefficient 1/2 is introduced for the convenience. For the longitudinal waves, $\mathbf{r}_l \equiv \mathbf{R}$, Eq. (28). The canonical equations for Eq. (35) are the standard ray equations of the geometrical optics or acoustics [14]

$$\frac{d\mathbf{p}_{lc}}{ds_l} = -\frac{\partial \mathcal{H}^l \left( \mathbf{p}_{lc}, \mathbf{r}_{lc} \right)}{\partial \mathbf{r}_l} = \frac{1}{2} \nabla n_l^2 \left( \mathbf{r}_{lc} \right) \ , \quad \frac{d\mathbf{r}_{lc}}{ds_l} = \frac{\partial \mathcal{H}^l \left( \mathbf{p}_{lc}, \mathbf{r}_{lc} \right)}{\partial \mathbf{p}_l} = \mathbf{p}_{lc} \ . \tag{36}$$

Here $s_l$ is the ray parameter connected with the ray length, $l$, as $dl = n_l ds_l$. Solutions of Eqs. (36), $\mathbf{p}_{lc}(s_l)$, $\mathbf{r}_{lc}(s_l)$, represent rays, i.e., the trajectories along which the centers of semiclassical wave packets move in the phase space $(\mathbf{p}_l, \mathbf{r}_l)$.

## 6. Evolution of transverse waves

The evolution of transverse waves can be regarded as that of optical phonons, i.e. particles with spin 1 and helicity $\sigma = \pm 1$.

**6.1. Basic equations.** Taking Eq. (4) into account, we rewrite equation (26) as

$$\hat{H}^t \psi = 0 \ , \quad \hat{H}^t (\mathbf{p}, \mathbf{R}) = p^2 - n_t^2 (\mathbf{R}) - \lambdabar_{t0} \nabla n_t^2 (\mathbf{R}) \hat{\mathbf{A}}^t (\mathbf{p}) - i \lambdabar_{t0} \nabla \ln \mu (\mathbf{R}) \mathbf{p} \ . \tag{26a}$$

Unlike the longitudinal waves, the transverse waves have extra degree of freedom, namely, the polarization which can be considered as the spin of optical phonons. Therefore, the geometrical acoustics ansatz takes the form $\psi = \mathbf{e}(\mathbf{R}) A_t (\mathbf{R}) \exp \left[ i \lambdabar_{t0}^{-1} \Phi_t (\mathbf{R}) \right]$, where the unit vector of polarization, $\mathbf{e} = \begin{pmatrix} e^+ \\ e^- \end{pmatrix}$, $\mathbf{e}^\dagger \mathbf{e} = 1$, is introduced. In the zero approximation in $\varepsilon$ (i.e. in $\lambdabar_{t0}$) we obtain the eikonal equation

$$\left( \nabla \Phi_t \right)^2 - n_t^2 = 0 \ , \tag{37}$$

whereas the terms of the first order in Eq. (26a) give rise to two equations describing the variations of the wave amplitude and the evolution of the polarization vector, respectively:

$$2 \nabla \Phi_t \nabla A_t + \left[ \nabla^2 \Phi_t + \nabla \Phi_t \nabla \ln \mu \right] A_t = 0 \ , \tag{38}$$

$$2i \left( \nabla \Phi_t \nabla \right) \mathbf{e} + \left( \nabla n_t^2 \hat{\mathcal{A}}^t \right) \mathbf{e} = 0 \ . \tag{39}$$

Here, similarly to the previous Section, $\mathbf{p}_t = \nabla \Phi_t = \lambdabar_{t0} \mathbf{k}_t$, and from now on the Berry gauge potential and field are considered in the $\mathbf{p}_t$-space: $\hat{\mathcal{A}}^t \equiv \hat{\mathbf{A}}^t (\mathbf{p}_t)$, $\hat{\mathcal{F}}^t \equiv \hat{\mathbf{F}}^t (\mathbf{p}_t)$, etc.

Eikonal equation (38) gives the dispersion equation $p_t = n_t$, and the transport equation (38) with Eq. (4) provides for the energy conservation law (the continuity equation) [3]:

$$\nabla\left(\rho A_t^2 \mathbf{c}_t\right) = 0 \ . \tag{40}$$

where $\mathbf{c}_t = c_t \mathbf{p}_t / p_t$ is the phase velocity vector.

**6.2. Polarization evolution and Berry phase.** Polarization evolution equation (39) and the ray equations are closely connected with each other. They represent the equations of motion for the translational and intrinsic (spin) degrees of freedom, respectively [9]. In the zero approximation in $\varepsilon$, the ray equations follow from the eikonal equation (37) and have the form completely similar to Eq. (36) [14]:

$$\dot{\mathbf{p}}_{tc}^{(0)} = \frac{1}{2}\nabla n_t^2\left(\mathbf{r}_{tc}^{(0)}\right), \quad \dot{\mathbf{r}}_{tc}^{(0)} = \mathbf{p}_{tc}^{(0)} \ . \tag{41}$$

From hereon dot stands for the derivative with respect to the ray parameter $s_t$: $dl = n_t ds_t$. Let us consider Eq. (39) on a zero-approximation ray, $\mathbf{p}_t = \mathbf{p}_{tc}^{(0)}(s_t)$, $\mathbf{r}_t = \mathbf{r}_{tc}^{(0)}(s_t)$. (We use here the zero approximation for rays, since the equation (39) has been derived from the terms of the order of $\varepsilon$.) Then, in Eq. (39) $(\nabla\Phi_t\nabla)\mathbf{e}(\mathbf{r}_{tc}^{(0)}) = \frac{d\mathbf{e}_c}{ds_t}$, $\mathbf{e}_c \equiv \mathbf{e}(\mathbf{r}_{tc}^{(0)})$, $\hat{\mathcal{A}}_c^t \equiv \hat{\mathcal{A}}^t(\mathbf{p}_{tc}^{(0)})$, and using the first equation (41) we obtain

$$\dot{\mathbf{e}}_c = i\left(\hat{\mathcal{A}}_c^t \dot{\mathbf{p}}_{tc}^{(0)}\right)\mathbf{e}_c \ . \tag{42}$$

Since $\hat{\mathcal{A}}^t$, Eq. (27), is an Abelian potential, equation (42) can be integrated,

$$\mathbf{e}_c = \exp\left[i\int_0^{s_t}\hat{\mathcal{A}}_c^t \dot{\mathbf{p}}_{tc}^{(0)} ds_t\right]\mathbf{e}_{c0} = \exp\left[i\int_C \hat{\mathcal{A}}^t(\mathbf{p}_t) d\mathbf{p}_t\right]\mathbf{e}_{c0} \ , \tag{43}$$

where $\mathbf{e}_{c0} \equiv \mathbf{e}_c(s_t = 0)$, and $C$ is the contour of the zero-approximation evolution in the momentum $\mathbf{p}_t$-space: $C = \{\mathbf{p}_t = \mathbf{p}_{tc}^{(0)}(s_t)\}$. Expression (43) shows that waves of the right-hand and left-hand circular polarizations acquire additional phases upon the evolution, that are equal in the absolute values but are of opposite signs: $\mathbf{e}_c = \exp[i\Theta^B \hat{\sigma}_3]\mathbf{e}_{c0}$, or

$$\begin{pmatrix} e_c^+ \\ e_c^- \end{pmatrix} = \begin{pmatrix} e^{i\Theta^B} e_{c0}^+ \\ e^{-i\Theta^B} e_{c0}^- \end{pmatrix} \ . \tag{44}$$

The phase $\Theta^B = \Theta^B(C) = \int_C \mathcal{A}^t d\mathbf{p}_t$ is the Berry geometrical phase, similar to that of light and is determined by the contour integral of the Berry gauge potential in $\mathbf{p}_t$-space. From Eq. (27) it follows (compare with [2]):

$$\Theta^B = \int_C \mathcal{A}^t d\mathbf{p}_t = \int_C \cos\vartheta \, d\varphi \ , \tag{45}$$

where $(p_t, \vartheta, \varphi)$ are the spherical coordinates in the $\mathbf{p}_t$-space.

If a cyclic evolution takes place in $\mathbf{p}_t$-space, i.e. the contour $C$ is a loop, the contour integral can be reduced to a surface one, and the Berry phase is determined by the flux of the Berry gauge field of the 'magnetic monopole', Eqs. (29), (29a):

$$\Theta^B = \oint_C \mathcal{A}^t d\mathbf{p}_t = \int_S \mathcal{F}^t d^2\mathbf{p}_t = \int_S \sin\vartheta \, d\vartheta d\varphi = -\Omega \ . \tag{46}$$

Here $S$ is a surface strained on the loop $C$ ($C = \partial S$), and $\Omega$ is the solid angle at which the surface is seen from the origin of $\mathbf{p}_t$-space.

It follows from Eq. (44) that, for an arbitrary elliptic polarization of the wave, the Berry phase causes the rotation of the polarization ellipse at the angle $-\Theta^B$. This rotation was described by Rytov [2,3] and was detected for electromagnetic waves in the experiments by Ross and by Tomita and Chiao [2]. The Berry phase is non-zero upon cyclic evolution if the ray is a non-flat curve (e.g., a helix). In fact, the polarization ellipse rotates in accordance to the Levi-Civita parallel transport law along the curved ray in 3D space [2].

Note that the quantity $\sigma_c = \mathbf{e}_c^\dagger \hat{\sigma}_3 \mathbf{e}_c = \left( \mathrm{e}_c^{+2} - \mathrm{e}_c^{-2} \right) \in (-1,1)$ is conserved upon the evolution of the polarization vector, Eq. (42):

$$\dot{\sigma}_c = \dot{\mathbf{e}}_c^\dagger \hat{\sigma}_3 \mathbf{e}_c + \mathbf{e}_c^\dagger \hat{\sigma}_3 \dot{\mathbf{e}}_c = -i \mathbf{e}_c^\dagger \left( \hat{\mathcal{A}}_c^t \dot{\boldsymbol{p}}_{tc}^{(0)} \right) \hat{\sigma}_3 \mathbf{e}_c + i \mathbf{e}_c^\dagger \hat{\sigma}_3 \left( \hat{\mathcal{A}}_c^t \dot{\boldsymbol{p}}_{tc}^{(0)} \right) \mathbf{e}_c = 0 \ . \tag{47}$$

This reflects the fact that helicity is an adiabatic invariant in the evolution of massless particles. It is natural to refer to $\sigma_c$ as the "mean helicity" or the "degree of helicity" in the center of the wave packet.

**6.3. Ray equations and topological spin transport of phonons.** Ray equations are the equations of characteristics (projected on the real phase space) of Eq. (26a). The last, imaginary term in Eq. (26a) does not contribute to these equations. However, the polarization term in Eq. (26a), (i.e. the term of the spin-orbit interaction of phonons, $\hat{H}_{SO}$) does contribute to the characteristics of the wave equation, despite its smallness of the order of $\varepsilon$. As a result, the rays of transverse acoustic waves are described by the Hamiltonian

$$\hat{\mathcal{H}}^t\left(\boldsymbol{p}_t, \mathbf{R}\right) = \frac{1}{2}\left[ p_t^2 - n_t^2(\mathbf{R}) - \lambdabar_{t0} \nabla n_t^2(\mathbf{R}) \hat{\mathcal{A}}^t(\boldsymbol{p}_t) \right]$$
$$\simeq \frac{1}{2}\left[ p_t^2 - n_t^2\left(\mathbf{R} + \lambdabar_{t0}\hat{\mathcal{A}}^t\right) \right] = \frac{1}{2}\left[ p_t^2 - n_t^2(\hat{r}_t) \right] = 0 \ , \tag{48}$$

where $\hat{r}_t = \mathbf{R} + \lambdabar_{t0}\hat{\mathcal{A}}^t$, Eq. (28). The canonical equations in usual coordinates $\mathbf{R}$,

$$\dot{\boldsymbol{p}}_{tc} = -\frac{\partial \hat{\mathcal{H}}^t(\dot{\boldsymbol{p}}_{tc}, \mathbf{R}_c)}{\partial \mathbf{R}} \ , \quad \dot{\mathbf{R}}_c = \frac{\partial \hat{\mathcal{H}}^t(\boldsymbol{p}_{tc}, \mathbf{R}_c)}{\partial \boldsymbol{p}_{tc}} \ ,$$

are not gauge invariant: their form depends on the choice of gauge for the potential $\hat{\mathcal{A}}^t$, and, hence, can not describe real rays. On the other hand, in the covariant coordinates $\hat{r}_t$, we get a gauge-invariant matrix-operator equations [6–12] which in the first approximation in $\varepsilon$ read

$$\dot{\hat{\boldsymbol{p}}}_{tc} = \frac{1}{2}\nabla n_t^2(\hat{r}_c) \ , \quad \dot{\hat{r}}_{tc} = \hat{\boldsymbol{p}}_{tc} + \lambdabar_{t0}\hat{\mathcal{F}}_c^t \times \dot{\boldsymbol{p}}_{tc}^{(0)} = \hat{\boldsymbol{p}}_{tc} - \lambdabar_{t0}\frac{\boldsymbol{p}_{tc}^{(0)} \times \dot{\boldsymbol{p}}_{tc}^{(0)}}{p_{tc}^{(0)3}}\hat{\sigma}_3 \ , \tag{49}$$

where $\hat{\mathcal{F}}_c^t \equiv \hat{\mathcal{F}}^t\left(\boldsymbol{p}_{tc}^{(0)}\right)$ and the $\varepsilon$-order term had been calculated on the zero-approximation ray. These equations are gauge-invariant with respect to $SU(2)$ gauge transformations related to the initial double degeneracy of the level and describe the trajectory of the wave packet center. Equations (49) can also be derived as a semiclassical limit of the Heisenberg quantum equations of motion for $\mathbf{p}$ and $\hat{r}_t$ [6,9,10,12] or from classical mechanics considerations [11]. In the former case, the polarization term related to the Berry gauge field appears due to the non-commutativity of coordinates $\hat{r}_t$, Eq. (31). Equations (49) are equations for matrix operators, and in order to find the real physical trajectories (rays) one has to make a quantum-mechanical convolution of operators with the polarization vector of the wave. In so doing, we obtain:

$$\dot{\boldsymbol{p}}_{tc} = \frac{1}{2}\nabla n_t^2(r_{tc}) \ , \quad \dot{r}_{tc} = \boldsymbol{p}_{tc} + \lambdabar_{t0}\mathcal{F}_c^t \times \dot{\boldsymbol{p}}_{tc}^{(0)} = \boldsymbol{p}_{tc} - \sigma_c \lambdabar_{t0}\frac{\boldsymbol{p}_{tc}^{(0)} \times \dot{\boldsymbol{p}}_{tc}^{(0)}}{p_{tc}^{(0)3}} \ , \tag{50}$$

where $(r_{tc})_i = \mathbf{e}_c^\dagger(\hat{r}_{tc})_i \mathbf{e}_c$, $(p_{tc})_i = \mathbf{e}_c^\dagger(\hat{p}_{tc})_i \mathbf{e}_c$, and $(\mathcal{F}_c^t)_i = \mathbf{e}_c^\dagger(\hat{\mathcal{F}}_c^t)_i \mathbf{e}_c$.

Equations (50) is one of the central results of the present paper. Analogous equations have been previously derived for the evolution of various quantum particles with spin: electrons, photons, quasiparticles in solids, etc. [6–12]. Equations (50) differ from the traditional ray equations of the geometrical optics and acoustics, Eqs. (41), by the additional polarization term proportional to the wavelength $\lambdabar_{t0}$. Since it contributes to the equation for the 'velocity' $\dot{\boldsymbol{r}}_{tc}$, it is frequently referred to as the "anomalous velocity" [12]. This term has the form of the 'Lorentz force' caused by the 'magnetic monopole' located at the origin of momentum space. Thus, the Berry gauge field reveals itself as being completely similar to the magnetic field but in momentum space rather than in coordinate space. In this respect, the Berry phase is an analogue of the Dirac phase (Aharonov–Bohm effect), whereas the additional term in the ray equations of motion is an analogue of the Lorentz force. It should be noted that the polarization term in Eq. (50) is directly connected to the Berry phase, Eqs. (43)–(46): it is the Berry phase that shifts the phase front of the wave and changes characteristics of the wave equation [8]. The reason why the additional term in Eq. (50) had been unnoticed in geometrical optics and acoustics for a long time, is that that the rays were associated with the characteristics of the eikonal (zero-approximation) equation (37), while the characteristics of the initial wave equation differ from them (in contrast to the case of longitudinal waves) already in the first order in $\varepsilon$.

The remarkable feature of the new term in Eqs. (50) is its dependence on the polarization of the wave. This means that the refraction of transverse waves becomes dependent on their polarization. In particular, the circularly polarized waves of opposite helicities shift in opposite directions orthogonally to the wave momentum. For quantum particles this phenomenon is treated as the appearance of the spin current, which is orthogonal to the direction of the particle motion and to the external applied force. Therefore, the effect is called the (intrinsic) spin Hall effect. Thus, equations (50) describe the *intrinsic spin Hall effect (or the topological spin transport) of optical phonons*.

Since the polarization correction in Eq. (50) is small, the perturbation method for rays [14] can be evoked. The first-order perturbations, $\boldsymbol{p}_{tc} - \boldsymbol{p}_{tc}^{(0)} = \delta\boldsymbol{p}_{tc}(s_t, \sigma_c)$, $\boldsymbol{r}_{tc} - \boldsymbol{r}_{tc}^{(0)} = \delta\boldsymbol{r}_{tc}(s_t, \sigma_c)$, obey equations

$$\delta\dot{\boldsymbol{p}}_{tc} = \frac{1}{2}(\delta\boldsymbol{r}_{tc}\nabla)\nabla n_t^2\left(\boldsymbol{r}_{tc}^{(0)}\right), \quad \delta\dot{\boldsymbol{r}}_{tc} = \delta\boldsymbol{p}_{tc} - \sigma_c\lambdabar_{t0}\frac{\boldsymbol{p}_{tc}^{(0)}\times\dot{\boldsymbol{p}}_{tc}^{(0)}}{p_{tc}^{(0)3}} . \qquad (51)$$

In the important special case $\delta\boldsymbol{p}_{tc} \equiv 0$, the second equation in (51) can be immediately integrated [8,10]. This leads to the expression for the deflection of ray, $\delta\boldsymbol{r}_{tc}$, in the form of a contour integral in the $\boldsymbol{p}_t$-space:

$$\delta\boldsymbol{r}_{tc} = -\sigma_c\lambdabar_{t0}\int_0^{s_t}\frac{\boldsymbol{p}_{tc}^{(0)}\times\dot{\boldsymbol{p}}_{tc}^{(0)}}{p_{tc}^{(0)3}}ds_t = -\sigma_c\lambdabar_{t0}\int_C\frac{\boldsymbol{p}_t\times d\boldsymbol{p}_t}{p_t^3} . \qquad (52)$$

Although the deflection (52) is proportional to the wavelength, it can be large because of its non-integrability and lead to observable phenomena (see optical examples in [7–10]). For closed trajectories in $\boldsymbol{p}_t$-space, the deflection (52) can be expressed by means of the Berry phase [8,10] as

$$\delta\boldsymbol{r}_{tc} = -\sigma_c\lambdabar_{t0}\frac{\partial\Theta^B}{\partial\boldsymbol{p}_{tc}^{(0)}} . \qquad (53)$$

As an example, the ray trajectories of transverse waves of right-hand and left-hand circular polarizations in a cylindrically symmetric waveguide medium are shown in Fig. 1. Since the optical and acoustic ray equations are identical, the calculations for the trajectories of rays in optical gradient-index waveguides [7,8] can be applied to the respective acoustic problem.

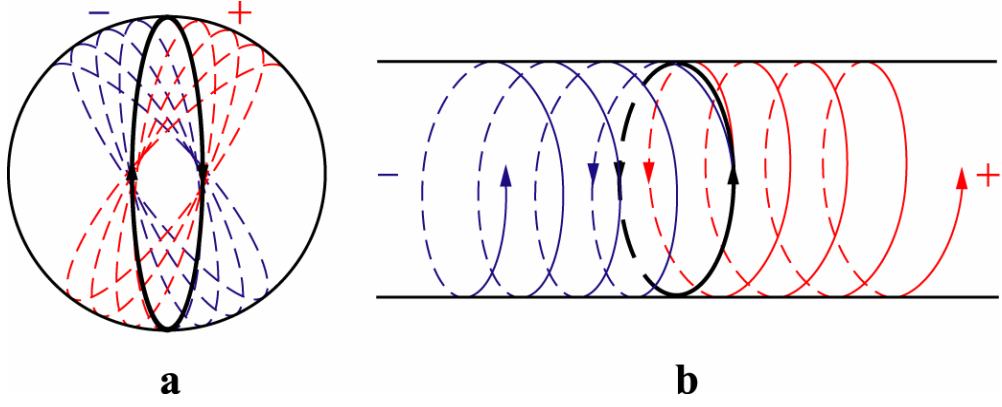

**Fig. 1.** Rays of righ-hand, «+», and left-hand, «−», circular polarizations in a waveguide smoothly inhomogeneous medium with cylindrical symmetry. Picture (a) shows the end view for the rays propagating along the waveguide axis, picture (b) displays the rays propagating across the waveguide and corresponding to the whispering gallery modes. The zero approximation rays are depicted by bold lines in both pictures.

### 7. Conservation of total angular moment of phonons.

It is important to note that the Berry gauge field, non-commutativity of coordinates, and the derived ray equations (50) are closely connected to the conservation of the total angular momentum of the transverse acoustic wave (optical phonons). The total angular momentum of phonon, which consists of the orbital and spin parts, can be written (in units $\lambdabar_{t0} = \hbar$) as [9]

$$\mathbf{j} = \mathbf{r}_{tc} \times \mathbf{p}_{tc} + \lambdabar_{t0}\sigma_c \frac{\mathbf{p}_{tc}}{p_{tc}} \ . \tag{54}$$

(We assume that the orbital angular momentum is determined only by the motion of the center of phonon, i.e., it does not carry any orbital angular momentum relative to the center [19].) Taking Eqs. (47) and (50) into account, the derivative of the total angular momentum along the ray equals

$$\dot{\mathbf{j}} = \dot{\mathbf{r}}_{tc} \times \mathbf{p}_{tc} + \mathbf{r}_{tc} \times \dot{\mathbf{p}}_{tc} + \lambdabar_{t0}\sigma_c \frac{\dot{\mathbf{p}}_{tc} p_{tc}^2 - \mathbf{p}_{tc}(\mathbf{p}_{tc}\dot{\mathbf{p}}_{tc})}{p_{tc}^3} = \mathbf{r}_{tc} \times \dot{\mathbf{p}}_{tc} = \frac{\mathbf{r}_{tc} \times \nabla n^2(\mathbf{r}_{tc})}{2} \ . \tag{55}$$

The first and the third terms in Eq. (55) are canceled due to the polarization term in the ray equations (50). From Eq. (55) it follows that in a spherically symmetric medium, $n_t(\mathbf{r}_t) = n_t(r_t)$, and $\nabla n^2(\mathbf{r}_{tc}) \parallel \mathbf{r}_{tc}$, the total angular momentum of the transverse acoustic wave is conserved and represents an integral of motion, $\dot{\mathbf{j}} = 0$. In a cylindrically symmetric medium, $n_t(\mathbf{r}_t) = n_t\left(\sqrt{x_t^2 + y_t^2}, z_t\right)$, the integral of motion is the $z$-component of the total angular momentum: $\dot{j}_z = 0$. Thus, it is the polarization term in the ray equations (50) that provides the conservation of the total angular momentum of the transverse waves. A detailed consideration of the connection between the angular momentum and the issues discussed (Berry phase, topological spin transport, and localizability) for photons can be found in [6,9,16,17].

## 8. Transverse Fedorov–Imbert shift

Another example of transverse polarization transport takes place in media with sharp inhomogeneities, which corresponds to the limit $\varepsilon \to \infty$. It is known in optics that a wave packet (or a beam) reflected or refracted from a flat interface between two homogeneous media experiences a small polarization-dependent transverse shift. This is called the Fedorov–Imbert shift, and it has been described theoretically and measured experimentally [20,21,18]. Analogously to the topological spin transport in smoothly inhomogeneous media, upon scattering at a sharp interface, the center of the wave packet is shifted from the plane of incidence, with the displacement proportional to the mean helicity of the incident wave. The Fedorov–Imbert shift also relates directly to the conservation of total angular momentum [18,21].

Here we consider the reflection of a monochromatic acoustic wave packet (or beam) from a plane boundary between an isotropic homogeneous medium and the vacuum. If the $z$ axis is orthogonal to the boundary, the $z$-component of the total angular momentum of waves, $\mathbf{J}$, is conserved, $J_z = \mathrm{const}$. When the wave packet is composed of $N$ phonons, its total angular momentum is given by $\mathbf{J} = N\mathbf{j}$, where $\mathbf{j}$ is defined by Eq. (54) for transverse waves and $\mathbf{j} = \frac{\bar{\lambda}_{t0}}{\bar{\lambda}_{l0}} \mathbf{r}_{lc} \times \mathbf{p}_{lc}$ for longitudinal wave packet that does not carry spin angular momentum. The reflection of acoustic waves represents two-channel scattering because there are two packets, transverse and longitudinal, in the reflected field [13]. If the energy reflection coefficients in the two channels (i.e. the number of phonons reflected in each one) are $\mathcal{R}_t$ and $\mathcal{R}_l$, respectively, ($\mathcal{R}_t + \mathcal{R}_l = 1$), then the conservation law, $J_{0z} = J_{tz} + J_{lz}$, yields [18]:

$$j_{0z} = \mathcal{R}_t j_{tz} + \mathcal{R}_l j_{lz} . \quad (56)$$

From here on the subscripts $0$, $t$, and $l$ denote quantities related to the incident, transverse-reflected, and longitudinal-reflected waves, respectively. If there are more than two channels in the system (for example, two reflected and two refracted waves), the conservation law for $J_z$ takes the form similar to Eq. (56) with the corresponding number of terms on the right-hand side. In the semiclassical approximation, when the characteristic dimensions of the wave packet are much larger than the wavelength, the reflection coefficients $\mathcal{R}_t$ and $\mathcal{R}_l$, are in fact the reflection coefficients of the central plane wave in the packet. Thus, when the problem of the reflection of a semiclassical wave packet is considered, the conservation law (56) contains only standard characteristics of the plane wave reflection, which can be easily calculated.

Unit central polarization vector $\mathbf{e}_c$ for the transverse wave can be parameterized by the single complex number $\chi$: $\mathbf{e}_c = \begin{pmatrix} 1-i\chi \\ 1+i\chi \end{pmatrix} / \sqrt{2(1+|\chi|^2)}$ [18] ($\chi$ is a ratio of the complex components of the displacement orthogonal to the plane of propagation and of the in-plane components). Then, $\sigma_c = \frac{2\,\mathrm{Im}\,\chi}{1+|\chi|^2}$ and the $z$-component of the total angular momentum of a single transverse phonon, Eq. (54), becomes:

$$j_z = -\Delta p_{cx} + \bar{\lambda}_{t0} \frac{2\,\mathrm{Im}\,\chi}{1+|\chi|^2} p_{cz} , \quad (57)$$

where $\Delta = y_c$ is the shift of the center of gravity of the wave packet along $y$ axis.

For a transverse elliptically polarized incident wave packet with central polarization $\chi$, using the acoustic Fresnel reflection coefficients [13], one can derive

$$\mathcal{R}_t = \frac{|R_1|^2 + |\chi_0|^2}{1+|\chi_0|^2} \ , \quad \mathcal{R}_l = \frac{c_l \cos\gamma_l}{c_t \cos\gamma_0}\frac{|R_2|^2}{1+|\chi_0|^2} \ , \quad \chi_t = \frac{\chi_0}{R} \ . \tag{58}$$

where $\chi_t$ characterizes the central polarization of the reflected transverse wave packet, $\gamma_0$ is the angle of incidence of the transverse wave, $\gamma_l$ is the angle of reflection of the longitudinal wave determined by the Snell's law (conservation law for the $x$-component of the momentum, $p_{cx} = \text{const}$): $\sin\gamma_l = \frac{c_l}{c_t}\sin\gamma_0$, and

$$R_1 = \frac{c_t^2 \sin 2\gamma_l \sin 2\gamma_0 - c_l^2 \cos^2 2\gamma_0}{c_t^2 \sin 2\gamma_l \sin 2\gamma_0 + c_l^2 \cos^2 2\gamma_0} \ , \quad R_2 = \frac{2 c_l c_t \sin 2\gamma_0 \cos 2\gamma_0}{c_t^2 \sin 2\gamma_l \sin 2\gamma_0 + c_l^2 \cos^2 2\gamma_0} \ , \tag{59}$$

are the Fresnel coefficients.

Substituting Eqs. (57)–(59) into Eq. (56), we obtain

$$\mathcal{R}_t \Delta_t + \mathcal{R}_l \Delta_l = -\lambdabar_{t0} \sigma_{c0} \cot\gamma_0 \left(1 + R_1\right) \ . \tag{60}$$

Here $\Delta_t$ and $\Delta_l$ are the transverse shifts of the reflected beams and $\Delta_0 = 0$. Equation (60) shows that at least one of the reflected wave packets does experience the transverse shift proportional to the helicity of the incident wave packet, $\sigma_{c0}$. Unfortunately, the single conservation law (56), (60) is not sufficient for the determination of two unknown shifts in the two-channel scattering [18]. To determine the explicit expressions for $\Delta_{t,l}$, one has to solve a complex problem of the reflection of the particular wave packet taking into account its spectral structure (see, e.g. [18] and paper by Nasalski in [20]).

It is worth noting that the ray equations in a smoothly inhomogeneous medium, Eqs. (50), can also be derived immediately from the expression for the transverse shift in the refraction on the interface between two media with a weak contrast of the refractive indices, $\delta n_t$. In such a case, the transverse wave is almost completely transformed into the refracted transverse wave (i.e. a one-channel scattering takes place) and the conservation of the normal component of the total angular momentum enables one to find small transverse shift of its center: $\Delta_t \approx \sigma_c \lambdabar_{t0} \frac{\delta n_t}{n_t} \tan\gamma_0$. The transition from small values to differentials in this problem gives the required equations (50) (see [7]). This fact emphasizes the common nature of the two polarization transport phenomena related to the opposite limits, $\varepsilon \to 0$ and $\varepsilon \to \infty$.

## 8. Conclusions

We have carried out a semiclassical analysis of the evolution of monochromatic linear acoustic waves in a smoothly-inhomogeneous isotropic medium. The modified geometrical acoustics has been developed, which accounts for the coupling between polarization and translational degrees of freedom of the transverse waves, i.e., the spin-orbit interaction of optical phonons. Similar to electrons, photons, etc., the spin-orbit interaction of phonons directly relates to the Berry gauge potential (connection) describing parallel transport in momentum space. The influence of the ray trajectories on the polarization brings about Berry phases of opposite signs for the right-hand and left-hand circularly polarized transverse waves and the Rytov rotation of the polarization ellipse. The reciprocal effect of the polarization influence on the ray trajectories is described by an additional term in the ray equations of motion, which has the form of the 'Lorentz force' caused by the Berry gauge field in momentum space. Because of this term, waves of different polarizations propagate along slightly different trajectories and waves of opposite helicities experience deflections in opposite directions, orthogonal to the ray direction and to the gradient of inhomogeneity ('external force'). It was shown that the polarization term makes the ray equations compatible with the conservation law of the total angular momentum of optical phonons. This conservation law also predicts the transverse polarization shift of the acoustic wave packet reflected from the flat sharp boundary. This is an acoustic analogue of the optical Fedorov–Imbert shift. The longitudinal acoustic waves contribute to the evolution of the transverse ones only when scattering by sharp inhomogeneities takes place; otherwise, in a smooth medium, the evolutions of two types of waves are independent.

The phenomena discussed above can manifest themselves in the following acoustic systems. First, the Berry phase observed as the Rytov rotation of the polarization plane reveals itself in the propagation of waves along helical trajectories, for instance, in helical rods of circular cross-section (i.e. helical acoustic waveguides), similar to the optical experiments of Ross and of Tomita and Chiao [2] (the possibility of such effect has also been mentioned by Segert [2]). The predicted transverse polarization deflection (spin Hall effect) of phonons is difficult to observe due to its smallness. However, it can be enhanced significantly, for instance, by the accumulation of deflections in circular waveguides [7,8] or in periodic media [10]. Polarization transport can also be noticeable in phononic crystals with additional inhomogeneity, similarly to photonic crystals [9]. Besides, the transverse topological transport can be dramatically increased when the beam carries additional (intrinsic) orbital angular momentum [22]. The spin Hall effect has been detected for photons [7,20,23] and recently for electrons in solids [24]. There is good reason to believe that in the near future polarization transport will also be measured in acoustics.

## Acknowledgement

The authors are grateful to V. Kulagin for fruitful discussions.

## Appendix A: Diagonalization transformation and gauge potentials via generators of $SO(3)$ group.

The generators of group $SO(3)$ are:

$$\hat{E}_1 = \begin{pmatrix} 0 & 0 & 0 \\ 0 & 0 & -i \\ 0 & i & 0 \end{pmatrix}, \quad \hat{E}_2 = \begin{pmatrix} 0 & 0 & i \\ 0 & 0 & 0 \\ -i & 0 & 0 \end{pmatrix}, \quad \hat{E}_3 = \begin{pmatrix} 0 & -i & 0 \\ i & 0 & 0 \\ 0 & 0 & 0 \end{pmatrix}. \quad (A1)$$

Operator $\exp\left[-i\alpha\hat{\mathbf{E}}\mathbf{n}\right]$, where $\hat{\mathbf{E}} \equiv (E_1, E_2, E_3)$ and $\mathbf{n} \in S^2 \subset \mathbb{R}^3$ is a unit vector, is a rotation about $\mathbf{n}$ axis by an angle $\alpha$. It can be calculated explicitly that

$$\left(\exp\left[-i\alpha\hat{\mathbf{E}}\mathbf{n}\right]\right)_{ij} = \delta_{ij}\cos\alpha - e_{ijk}n_k\sin\alpha + n_i n_j (1-\cos\alpha) \quad (A2)$$

($e_{ijk}$ is the unit antisymmetric tensor). Using (A2) one can show that the diagonalization transformation (11) is a rotation which can be presented as a combination of two rotations about $x$ and $z$ axes:

$$\hat{U} = \exp\left[-i\left(\phi - \frac{\pi}{2}\right)\hat{E}_3\right]\exp\left[i\theta\hat{E}_1\right]. \quad (A3)$$

Obviously, the diagonalization transformation is defined up to an arbitrary rotation in the plane orthogonal to $\mathbf{p}$. For the transformation (A3), a subsequent rotation about $z$ axis (which is directed along $\mathbf{p}$ after transformation (11) or (A3)) will not affect the diagonalization, i.e. the diagonalization scheme has the $SO(2) \cong U(1)$ degree of freedom $\hat{U} \to \hat{U}\exp\left[-i\alpha \hat{E}_3\right]$. Indeed,

$$\exp\left[-i\alpha \hat{E}_3\right] = \begin{pmatrix} \cos\alpha & -\sin\alpha & 0 \\ \sin\alpha & \cos\alpha & 0 \\ 0 & 0 & 1 \end{pmatrix}$$ and it does not mix up 'transverse' and 'longitudinal' sectors

in the diagonalized wave equation.

Pure gauge potential (16) induced by the rotational transformation (11) can also be represented by means of the generators (A1) as

$$\hat{A}_p = 0 \ , \ \hat{A}_\theta = -\frac{i}{p}\hat{E}_1 \ , \ \hat{A}_\phi = \frac{i}{p}\left[\cot\theta \, \hat{E}_3 - \hat{E}_2\right] . \tag{A4}$$

After neglecting the cross terms in this potential, i.e. after transition from Eq. (21) to Eqs. (22) and (23), only the term proportional to $\hat{E}_3$ survives in (A4) and gives rise to the Berry gauge potential (24). It is the upper left $2\times 2$ sector of $\hat{E}_3$ that equals Pauli matrix $\hat{\sigma}_2$ in Eq. (24). This shows explicitly that the mentioned degree of freedom of the rotations about $\mathbf{p}$ ($z$ axis) brings about $U(1)$ Berry connection originated from the generator $\hat{E}$ in the induced pure gauge potential and describing the Levi–Civita parallel transport along the ray [2].

## Appendix B: Appearance of pure gauge potential, Eqs. (13) and (14)

To derive Eqs. (13), (14), let us prove the following equality for operators acting on functions of a single coordinate $x$

$$U^{-1}(x)f\left(\frac{d}{dx}\right)U(x) = f\left(\frac{d}{dx} + U^{-1}(x)\frac{dU(x)}{dx}\right). \tag{B1}$$

Here $f$ is an analytical function, and operator $f\left(\frac{d}{dx}\right)$ is assumed as its Taylor series. We first consider $f\left(\frac{d}{dx}\right) = \left(\frac{d}{dx}\right)^l$, $l \in \mathbb{N}$. For $l=1$ the expression is true: $U^{-1}\frac{d}{dx}U = \frac{d}{dx} + U^{-1}\frac{dU}{dx}$. If equation (B1) is true for some $l = j$:

$$U^{-1}\left(\frac{d}{dx}\right)^j U = \left(\frac{d}{dx} + U^{-1}\frac{dU}{dx}\right)^j , \tag{B2}$$

then, for $l = j+1$ one has

$$U^{-1}\left(\frac{d}{dx}\right)^{j+1} U = U^{-1}\left(\frac{d}{dx}\right)^j \frac{d}{dx}U = U^{-1}\left(\frac{d}{dx}\right)^j U\frac{d}{dx} + U^{-1}\left(\frac{d}{dx}\right)^j \frac{dU}{dx} =$$
$$= U^{-1}\left(\frac{d}{dx}\right)^j U\frac{d}{dx} + U^{-1}\left(\frac{d}{dx}\right)^j UU^{-1}\frac{dU}{dx} = U^{-1}\left(\frac{d}{dx}\right)^j U\left(\frac{d}{dx} + U^{-1}\frac{dU}{dx}\right) = \tag{B3}$$
$$= \left(\frac{d}{dx} + U^{-1}\frac{dU}{dx}\right)^{j+1} ,$$

where we have used Eq. (B2). Thus, by induction, the equality (B1) is proven for any power function $f$, and, hence, for any analytic one. Equations (13), (14) represent a three-dimensional generalization of the equality (B1), which can be proven in a similar way.

# References

1. B.A. Auld, *Acoustic fields and waves in solids* (John Wiley & Sons, New York, 1973; Krieger Pub Co, 1990).
2. For review see S.I. Vinnitski et al., Usp. Fiz. Nauk **160**(6), 1 (1990) [Sov. Phys. Usp. **33**, 403 (1990)]; *Topological Phases in Quantum Theory*, edited by B. Markovski and S.I. Vinitsky (World Scientific, Singapore, 1989); see also original papers S.M. Rytov, Dokl. Akad. Nauk. SSSR **18**, 263 (1938) [reprinted in the above book]; V.V. Vladimirskii, Dokl. Akad. Nauk. SSSR **31**, 222 (1941) [reprinted in the above book]; J.N. Ross, Opt. Quantum Electron. **16**, 455 (1984); A. Tomita and R.Y. Chiao, Phys. Rev. Lett. **57**, 937 (1986); R.Y. Chiao and Y.S. Wu, Phys. Rev. Lett. **57**, 933 (1986); F.D.M. Haldane, Opt. Lett. **11**, 730 (1986); M.V. Berry, Nature **326**, 277 (1987); J. Segert, Phys. Rev. A **36**, 10 (1987).
3. M.L. Levin and S.M. Rytov, Akustich. Zhurn. **2**, 173 (1956); F.C. Karal and J.B. Keller, J. Acoust. Soc. Am. **31**, 694 (1958).
4. R.G. Littlejohn and W.G. Flynn, Phys. Rev. A **44**, 5239 (1991).
5. H. Mathur, Phys. Rev. Lett. **67**, 3325 (1991); J. Bolte and S. Keppeler, Ann. Phys. (New York) **274**, 125 (1999).
6. A. Bérard and H. Mohrbach, Phys. Lett. A **352**, 190 (2006), hep-th/0404165; K.Yu. Bliokh, Europhys. Lett. **72**, 7 (2005).
7. V.S. Liberman and B.Ya. Zel'dovich, Phys. Rev. A **46**, 5199 (1992).
8. K.Yu. Bliokh and Yu.P. Bliokh, Phys. Rev. E **70**, 026605 (2004); Phys. Lett. A **333**, 181 (2004); physics/0402110.
9. M. Onoda, S. Murakami, and N. Nagaosa, Phys. Rev. Lett. **93**, 083901 (2004).
10. K.Yu. Bliokh and V.D. Freilikher, Phys. Rev. B **72**, 035108 (2005).
11. C. Duval, Z. Horváth, and P.A. Horváthy, Phys. Rev. D **74**, 021701 (2006); math-ph/0509031 (unpublished).
12. D. Culcer, A. MacDonald, and Q. Niu, Phys. Rev. B **68**, 045327 (2003); S. Murakami, N. Nagaosa, and S.-C. Zhang, Science **301**, 1348 (2003); J. Sinova et al., Phys. Rev. Lett. **92**, 126603 (2004); see also G. Sundaram and Q. Niu, Phys. Rev. B **59**, 14915 (1999) and reviews K.Yu. Bliokh and Yu.P. Bliokh, Ann. Phys. (New York) **319**, 13 (2005); S. Murakami, Adv. Solid State Phys. **45**, 197 (2005).
13. L.D. Landau and E.M. Lifshits, *Theory of elasticity* (Pergamon Press, Oxford, 1986).
14. Yu.A. Kravtsov and Yu.I. Orlov, *Geometrical Optics of Inhomogeneous Medium* (Springer-Verlag, Berlin, 1990); V. Červený, *Seismic Ray Theory* (Cambridge University Press, 2001).
15. Actually, the cross terms can be reduced by a one more transformation of $\tilde{u}$ close to the unit one. Such transformation will account for the small changes in the polarization of eigen modes, and make no contribution to the Hamiltonian and the equations of motion in the approximation under consideration.
16. B.-S. Skagestam, hep-th/9210054.
17. I. Bialynicki-Birula and Z. Bialynicki-Birula, Phys. Rev. D **35**, 2383 (1987).
18. K.Yu. Bliokh and Yu.P. Bliokh, Phys. Rev. Lett. **96**, 073903 (2006).
19. Beams or single photons with optical vortices (e.g., Laguerre–Gaussian beams) carry additional intrinsic angular momentum, see *Optical Angular Momentum*, edited by L. Allen, S.M. Barnett, and M.J. Padgett (Taylor & Francis, 2003).
20. F.I. Fedorov, Dokl. Akad. Nauk SSSR **105**, 465 (1955); C. Imbert, Phys. Rev. D **5**, 787 (1972); O. Costa de Beauregard, Phys. Rev. **139**, B1443 (1965); H. Schilling, Ann. Phys. (Berlin) **16**, 122 (1965); J. Ricard, Nouv. Rev. Opt. Appl. **1**, 273 (1970); D.G. Boulware, Phys. Rev. D **7**, 2375 (1973); N. Ashby and S.C. Miller, Jr., Phys. Rev. D **7**, 2383 (1973);

N.N. Punko and V.V. Filippov, Pis'ma v ZhETF **39**, 18 (1984) [JETP Lett. **39**, 20 (1984)]; L. Dutriaux, A. Le Floch, and F. Bretenaker, Europhys. Lett. **24**, 345 (1993); W. Nasalski, J. Opt. Soc. Am. A 13, 172 (1996); F. Pillon, H. Gilles, and S. Girard, Appl. Opt. **43**, 1863 (2004).

21. M.A. Player, J. Phys. A: Math. Gen. **20**, 3667 (1987); V.G. Fedoseyev, *ibid.* **21**, 2045 (1988); W.N. Hugrass, J. Mod. Opt. **37**, 339 (1990).
22. K.Yu. Bliokh, Phys. Rev. Lett. **97**, 043901 (2006); V.G. Fedoseev, Opt. Comm. **193**, 9 (2001); R. Dasgupta and P.K. Gupta, Opt. Comm. **257**, 91 (2006).
23. A.V. Dooghin et al., Phys.Rev. A **45**, 8204 (1992).
24. W.-L. Lee et al., Science **303**, 1647 (2004); J. Wunderlich et al., Phys. Rev. Lett. **94**, 047204 (2005); Y.K. Kato et al., Science **306**, 1910 (2004); B.A. Bernevig and S.-C. Zhang, cond-mat/0412550.